\documentclass[epj]{svjour}

\usepackage{graphicx}
\usepackage{amsmath}
\usepackage{amssymb}

\newcommand{\be}{\begin{equation}}
\newcommand{\ee}{\end{equation}}

\title{Casimir Force between Atomically Thin Gold Films}

\date{\today}

\author{M. Bostr{\"o}m\inst{1} \and C. Persson\inst{1,2} \and Bo E. Sernelius\inst{3}\thanks{E-mail: bos@ifm.liu.se} }

\institute{
	\inst{1} Department of Materials Science and Engineering, Royal Institute of Technology, SE-100 44 Stockholm, Sweden, EU\\
	\inst{2} Department of Physics, University of Oslo, P.O.  Box 1048 Blindern, N-0316 Oslo, Norway\\
	\inst{3} Division of Theory and Modeling, Department of Physics,  Chemistry and Biology, Link\"{o}ping University, SE-581 83 Link\"{o}ping, Sweden, EU }

\begin{document}

\abstract{
We have used density functional theory  to calculate the anisotropic dielectric functions for ultrathin gold sheets (composed of 1, 3, 6, and 15 atomic layers). Such films are important components  in nano-electromechanical systems. When using correct dielectric functions rather than bulk gold dielectric functions  we predict an enhanced attractive Casimir-Lifshitz force (at most around 20\%) between two atomically thin gold sheets.  For thicker sheets the dielectric properties and the corresponding Casimir forces approach those of gold half-spaces. The magnitude of the corrections that we predict  should, within the today's level of accuracy in Casimir force measurements, be  clearly detectable.
}


\maketitle

\section{Introduction}

 Casimir\,\cite{Casi} predicted already  in 1948  that boundary effects on the electromagnetic fluctuations can produce attraction between a pair of parallel, closely spaced, perfectly conducting plates. This work was later extended to real materials by Lifshitz\,\cite{Lifshitz,Dzya}, and  by Parsegian and Ninham\,\cite{NinPar1}. Since the famous experiments of Deryaguin and Abrikossova\,\cite{Der} there has been much interest in phenomena that measure the van der Waals forces acting between macroscopic bodies. The early experiments that measured the forces between quartz and metal plates covered only the retarded region. The experiments of Tabor and Winterton\,\cite{Tab} and subsequently of Israelachvili and Tabor\,\cite{IsraTabor,White} fitted the potential to a power law of $1/d^{n}$ ($d$ being the distance) where $n$ varied from non retarded ($n = 3$) to fully retarded ($n = 4$) value. There was a gradual transition from non-retarded  to retarded forces as the separation was increased from {120 \AA}  to {1300 \AA}\,\cite{IsraTabor}.  Lamoreaux performed the first high accuracy measurement\,\cite{Lamo} of Casimir forces between metal surfaces in vacuum\,\cite{Dzya,Bost2}.  The first measurements of Casimir-Lifshitz forces applied to micro-electromechanical systems were performed by Chan {\it et al.}\,\cite{ChanScience,ChanPRL} and somewhat later by Decca {\it et al.}\,\cite{Decca}. Several new very recent high precision measurements have been performed\,\cite{Decca1,Garcia,Sush,Umar}.  An interesting aspect of the Casimir-Lifshitz force is that according to theory it can  be either attractive or repulsive\,\cite{Dzya,Tas,Rich1,Rich,bosserPRA2012,bostAPL2012,Phan11}.
Casimir-Lifshitz repulsion was measured four decades ago for films of liquid helium (10-200 \AA) on smooth surfaces\,\cite{AndSab}.  Only a few direct force measurements of repulsive Casimir-Lifshitz forces have been reported in the literature\,\cite{Mund,Milling,Lee,Feiler,Zwol1}.

Surfaces that are very close interact in vacuum with a van der Waals force\,\cite{Casi,Dzya,Der,Tab,Maha,Lond,Ser,Milt,Pars,Ninhb,Rodriguez}.
The van der Waals force between thin isotropic metal films follows a  fractional power law related to the two dimensional plasmon dispersion\,\cite{Barash1,Barash2,BostromPRB,Das,Pincus,Tas}. As the separation between the surfaces increases the interaction takes a weaker (retarded) Casimir form\,\cite{Casi,Dzya}. As demonstrated by Benassi and Calandra\,\cite{Benassi} the Casimir force between ultrathin (1 to 10 nm thick) conducting films depends on the anisotropy of the films.
In the recent Casimir experiments the metal films coating the objects were thick enough for the objects to be considered as bulk metal objects. The metal films were assumed to be homogeneous and isotropic. This is not always so in reality.  Svetovoy {\it et al.}\,\cite{Svetovoy2008}  measured both the optical properties and Casimir forces for 1000-4000 {\AA} thick gold films on substrates. 
As a reference curve for the dielectric function of thin films they\,\cite{Svetovoy2008} chose one  which was calculated with handbook data\,\cite{Palik}.
The Casimir force evaluated for their films was considerably weaker compared to that calculated with the reference curve ($5\%$  to $14 \%$ weaker at a distance of 100 nm between the films). Thus, the dielectric function and force are sensitive to how the films are prepared. Thin films of gold can even be insulating\,\cite{Wal} due to disorder. In Reference\,\cite{Esq} one calculated the force near the insulator-conductor transition in thin gold films. In the present work we are not concerned with these effects. We assume homogeneous ordered metallic films. We address what happens when the metal films are extremely thin.

We first present the theory used to calculate anisotropic dielectric functions of atomically thin gold films. Such films are important components in nano-electromechanical systems (NEMS).  Then we demonstrate that there is an enhancement of the Casimir-Lifshitz force (of the order of 20 \%)  when proper dielectric functions for ultrathin films are used rather than the dielectric function appropriate for bulk gold. As the film thickness increases the use of isotropic bulk dielectric function becomes an increasingly good approximation. The long range entropic Casimir asymptote originates entirely from  the transverse magnetic zero frequency mode in agreement with recent experiments\,\cite{Garcia}. We end with a brief summary and an outlook.

\section{Calculation of dielectric permittivities of atomically thin gold films}
\begin{figure}
\includegraphics[width=8.5cm]{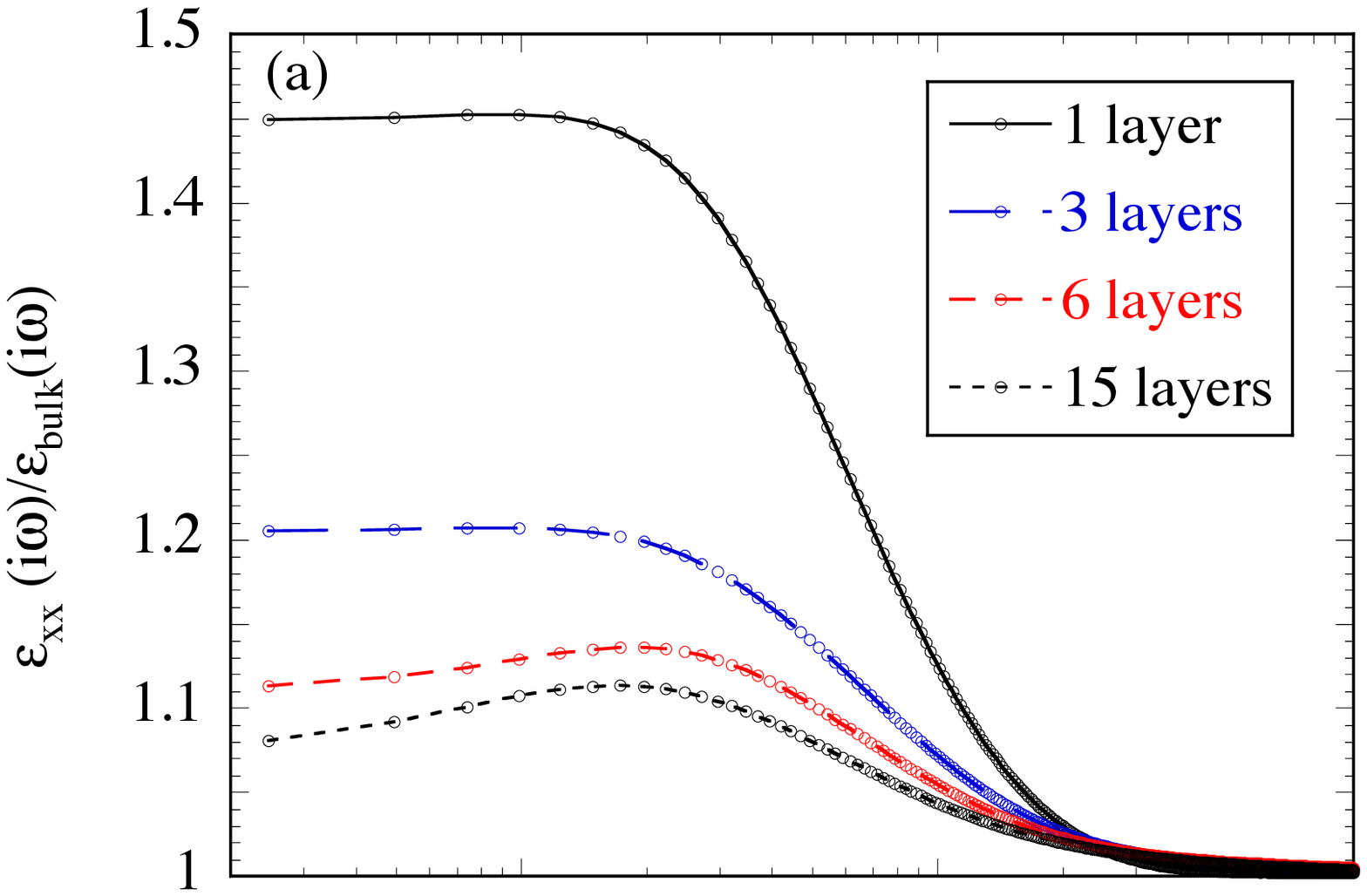}
\includegraphics[width=8.5cm]{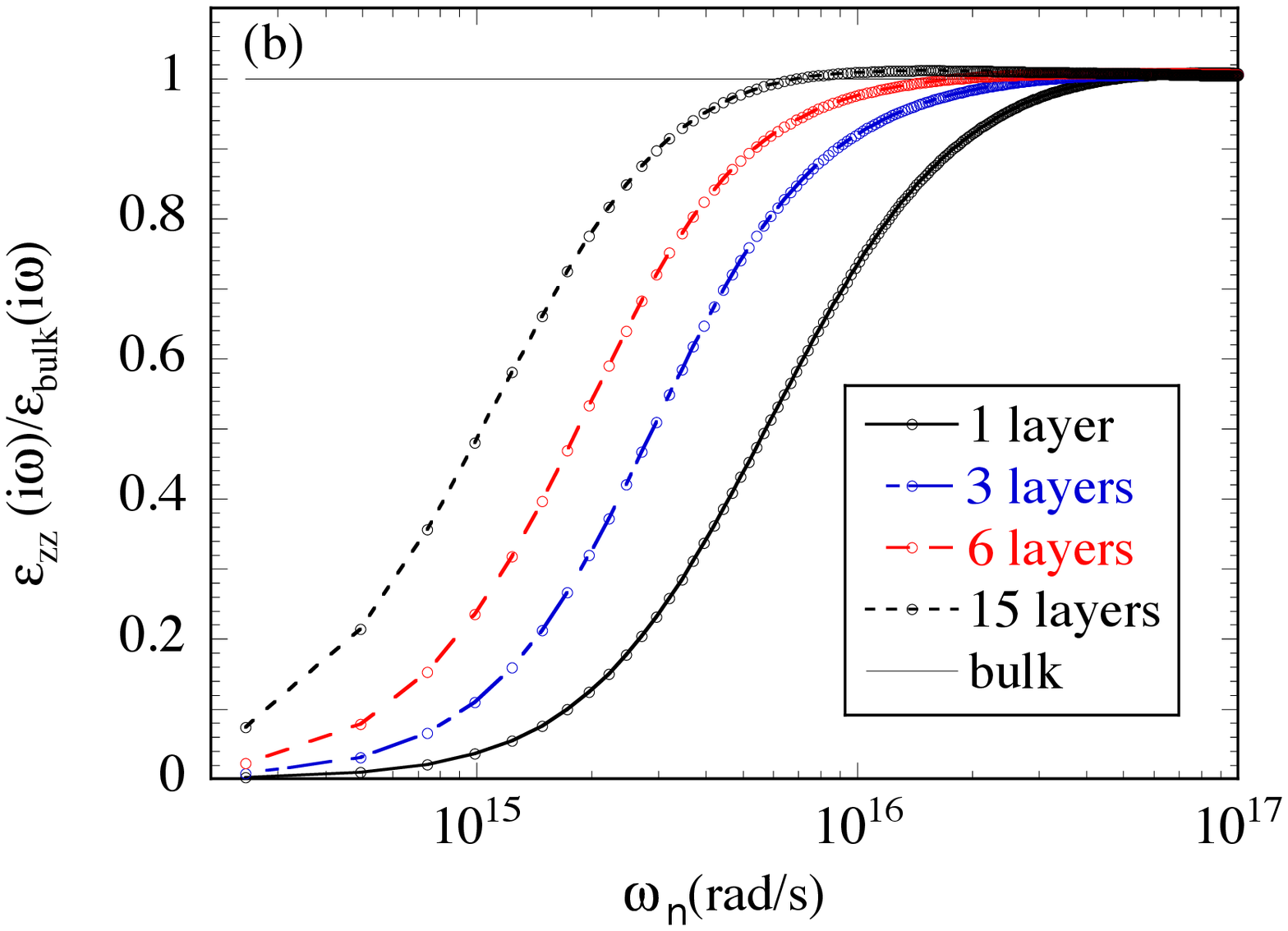}
\caption{ (Color online) Ratio between the diagonal elements of the dielectric tensor of  ultrathin gold sheets and that of bulk gold. The in-plane elements, $\varepsilon_{xx}=\varepsilon_{yy}$,  (a),  are enhanced compared to the bulk value while the perpendicular component $\varepsilon_{zz}$, (b), is reduced in value. $\varepsilon_{zz}(i\omega_1) =$ 6.4, 21.9, 60.2, and 198 for sheets composed of $N = 1$, 3, 6, and 15 atomic layers, respectively, whereas bulk gold has a value of $\varepsilon(i\omega_1) = 2700$. The lowest nonzero frequency $\omega_1$ is $ \approx 2.47 \times {10^{14}}$ rad/s.}
\label{figu1}
\end{figure}

We have calculated the  anisotropic dielectric function of ultrathin gold sheets  (1, 3, 6, and 15 atomic layers) plus the corresponding results for thick gold plates. The calculations of the dielectric function were performed within the density
functional theory, employing the local density approximation in conjunction
with the projector augmented wave method\,\cite{Ahjua,Gajdos,Kresse}. We modeled the layer
structures by supercells oriented in the crystalline (111) direction.
The supercells have hexagonal symmetry with a height of 56.5 \AA   (i.e.,  24 atomic layers).
The imaginary part of the dielectric function was calculated in the long wave
length limit from the optical momentum matrix, and the intraband contribution
was included by the Drude-like model with a damping of $\gamma$ = 5 meV. The dielectric function on the imaginary
frequency axis was determined from the Kramers-Kronig dispersion relation.
 The ratio of these dielectric functions to the corresponding dielectric function of bulk gold are shown in Figure\,\ref{figu1} (where $z$ is the (111) direction perpendicular to the thin film and $x$ is a direction within the film). These dielectric functions have been used together with the expression for the anisotropic Lifshitz force given by Benassi and Calandra\,\cite{Benassi},  with proper extension to include finite temperature effects\,\cite{Maha}.  While Casimir forces  between thin films (e.g. lipid films in water) have been known for more than 40 years\,\cite{Maha,Ninhb,Parsegian1971} they have not been explored for the present case of ultrathin anisotropic conducting films.  Nanotechnological advances and refined  atomic models of ultathin gold sheets now allow their exploitation.

\begin{figure}
\includegraphics[width=8.5cm]{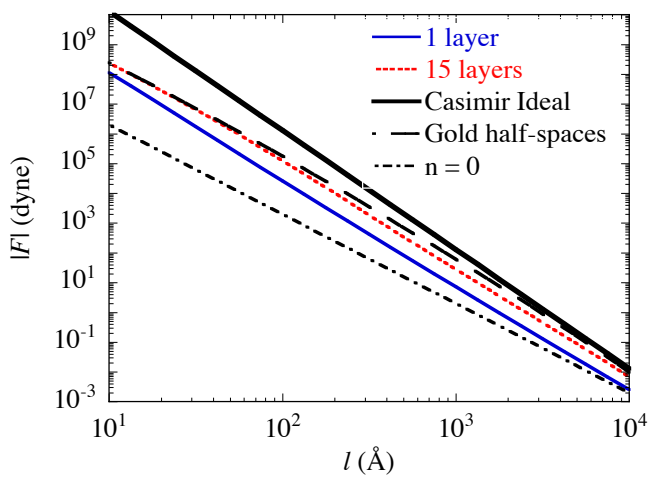}
\caption{(Color online) Casimir-Lifshitz force between ultrathin gold films, between gold half-spaces, and between ideal (perfectly reflecting) metal surfaces. For comparison we also show the $n=0$ entropic long range asymptotic force, which is the same for all cases except for the Casimir ideal case.}
\label{figu2}
\end{figure}

\section{Theory and Numerical Results}

The simlest way to find retarded van der Waals or Casimir-Lifshitz force is in terms of the electromagnetic normal modes\,\cite{Maha,Ser,Benassi} of the system. 
At finite temperature the force ($F$) originates from changes in the Helmholtz' free energy and can be written as

\begin{equation}
\begin{array}{l}
F = \frac{{ - {k_B}T}}{\pi }\int_0^\infty  k dk\sum\limits_{n = 0}^\infty  {^{'}\gamma \left( {i{\omega _n}} \right)\left[ {\frac{{{Q_{TM}}{{(i{\omega _n})}^2}}}{{1 - {Q_{TM}}{{(i{\omega _n})}^2}}}} \right.} \\
\quad \quad \quad \quad \quad \quad \quad \quad \quad \quad \quad \quad \quad \quad  + \left. {\frac{{{Q_{TE}}{{(i{\omega _n})}^2}}}{{1 - {Q_{TE}}{{(i{\omega _n})}^2}}}} \right],
\end{array}
\label{equ1}
\end{equation}
where

\begin{equation}
{Q_{TM/TE}} = \frac{{{\rho _{TM/TE}}(1 - {e^{ - 2D{\gamma _{TM/TE}}}})}}{{1 - \rho_{TM/TE}^2{e^{ - 2D{\gamma_{TM/TE}}}}}}{e^{ - \gamma l}}.
\label{equ2}
\end{equation}
Here,

\begin{equation}
{\rho _{TM}} = \frac{{{\gamma _{TM}}(i{\omega _n}) - \gamma (i{\omega _n}){\varepsilon _{xx}}(i{\omega _n})}}{{{\gamma _{TM}}(i{\omega _n}) + \gamma (i{\omega _n}){\varepsilon _{xx}}(i{\omega _n})}},
\label{equ3}
\end{equation}

\begin{equation}
{\rho _{TE}} = \frac{{{\gamma _{TE}}(i{\omega _n}) - \gamma (i{\omega _n})}}{{{\gamma _{TE}}(i{\omega _n}) + \gamma (i{\omega _n})}},
\label{equ4}
\end{equation}

\begin{equation}
\gamma (i \omega_n) =\sqrt{k^2+\omega^2/c^2},
\label{equ5}
\end{equation}

\begin{equation}
\gamma_{TE} (i \omega_n) =\sqrt{k^2+\varepsilon_{xx} \omega^2/c^2},
\label{equ6}
\end{equation}
and

\begin{equation}
\gamma_{TM} (i \omega_n) =\sqrt{\varepsilon_{xx} [(k^2/\varepsilon_{zz})+\omega^2/c^2]}.
\label{equ7}
\end{equation}

In Equation\,(\ref{equ2}) $l$ is the distance between the surfaces of the two sheets and $D=[2 d_0 + (N-1) d]$ is the film thickness ( {$d_0=d/2$},  {$d=2.35 $\AA} and $N$ is the number of atomic layers). With this definition the thickness of a single layer is the same as the distance between two atomic layers. The integral over frequency appropriate for zero temperature\,\cite{Benassi} has been replaced by a summation over discrete Matsubara frequencies to account for finite temperature. The prime on the summation sign indicates that the $n = 0$ term should be divided by two. 
For planar structures the quantum number that characterizes the normal modes is  $\bf k$, the wave vector in the plane of the interfaces, and there are two mode types, transverse magnetic (TM) and transverse electric (TE).

We show in Figure\,\ref{figu2} the Casimir-Lifshitz force between atomically thin gold sheets (exemplified with the cases of 1 layer and 15 layers). All results are for $T=300$K. The result is compared with the Casimir-Lifshitz force between two gold half-spaces and between two ideal metal surfaces. Included is also the long-range entropic contribution ($n=0$ term in the Matsubara summation).  This is the same for all cases except for the Casimir ideal case where this contribution is twice as big. For the range covered in the figure temperature effects have not yet become important. The force between two gold half-spaces is weaker than the force between two ideal metal half-spaces.  It approaches this force at the rightmost end of the figure where retardation effects prevent the crossing of the curves. The force between two 15-layer films approaches the force between two gold half-spaces well before the leftmost end of the figure is reached. For the 1-layer films this happens much later, outside the figure.

\begin{figure}
\includegraphics[width=8.5cm]{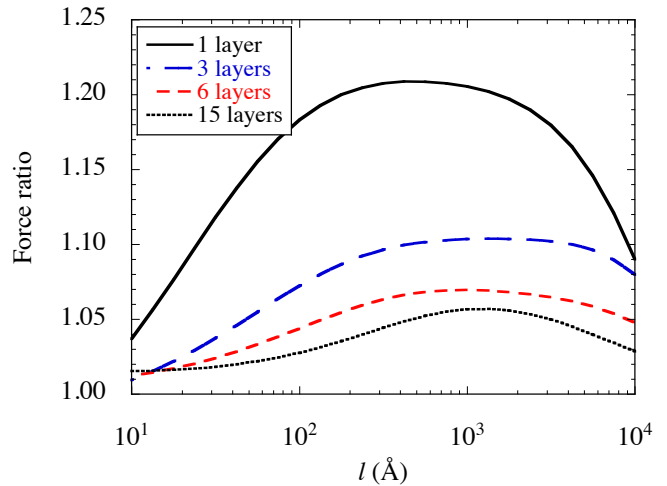}
\caption{(Color online) Ratio of the Casimir-Lifshitz force between thin films calculated with realistic anisotropic dielectric functions and from the corresponding force calculated using the isotropic bulk dielectric function.}
\label{figu3}
\end{figure}
\begin{figure}
\includegraphics[width=8.5cm]{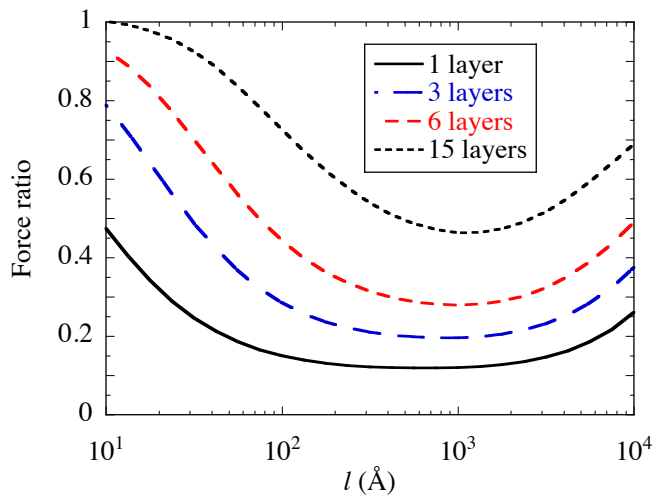}
\caption{(Color online) Ratio of Casimir-Lifshitz force between thin films and corresponding force between gold half-spaces.}
\label{figu4}
\end{figure}

In order to study the effect of using dielectric functions calculated for thin films we present in  Figure\,\ref{figu3} the ratio between Casimir-Lifshitz force calculated with realistic anisotropic dielectric functions to the corresponding calculation using isotropic bulk dielectric function.  We observe corrections in the range between 5 and 20 \% in the range most accessible to force measurements.

We finally demonstrate in  Figure\,\ref{figu4}. how the Casimir-Lifshitz force between thin gold films is reduced compared to the corresponding force between gold half-spaces.

To be noted is that the maximum deviation from unity in both Figures\,\ref{figu3} and \ref{figu4} occurs in the middle portion of the figures. For large separations only small frequencies contribute. For zero frequency ${\rho _{TM}} =  - 1$ and the factors containing the film thickness in Equation\,(\ref{equ2}) cancel. Furthermore ${\rho _{TE}} = 0$ and ${Q_{TE}} = 0$ for zero frequency. This behavior results from the properties of the low frequency screening: ${\varepsilon _{xx}}\left( \omega  \right)$ goes towards infinity at low frequencies and ${\omega ^2}{\varepsilon _{xx}}\left( \omega  \right)$ goes towards zero. For small separations the force between two films approaches the force between two half-spaces\,\cite{BostromPRB} and the anisotropy effects are reduced.

\section{Conclusions}

It has been stressed\,\cite{Zwol1} how crucial it is in Casimir force calculations to use accurate dielectric functions from optical data or from calculations. We demonstrate in this article that when performing calculations on Casimir-Lifshitz forces between ultrathin conducting films it is not sufficient to use dielectric functions relevant for thick films or half-spaces. Rather we have demonstrated the importance of using proper anisotropic dielectric functions from density functional calculations when performing calculations of Casimir forces between atomically thin gold sheets. These ultrathin films have interesting applications in nano-technology, including use in NEMS. Addition of ultrathin metallic coatings to surfaces in solution have for instance been predicted to create quantum levitation of  NEMS\,\cite{bostAPL2012}.  It remains to be seen how such predictions may be influenced when using accurate dielectric functions of atomically thin metal films.

\begin{acknowledgement}
 C.P. and M. B. acknowledge support from VR (Contract No. 90499401
) and STEM (Contract No. 34138-1).   B.E.S. acknowledges financial support from VR (Contract No. 70529001). 
\end{acknowledgement}

\end{document}